\documentclass[preprint,floatfix] {revtex4} 
\newcommand{\rvec}{\mathrm {\mathbf {r}}} 
 
\newcommand{\Rvec}{\mathrm {\mathbf {R}}} 
\newcommand{\Cvec}{\mathrm {\mathbf {C}}} 
\newcommand{\Svec}{\mathrm {\mathbf {S}}} 
\newcommand{\Ivec}{\mathrm {\mathbf {I}}} 
\newcommand{\kvec}{\mathrm {\mathbf {k}}} 
\newcommand{\Fvec}{\mathrm {\mathbf {F}}} 
\newcommand{\Pvec}{\mathrm {\mathbf {P}}} 

\newenvironment{rcases}
  {\left.\begin{aligned}}
  {\end{aligned}\right\rbrace}
  
\usepackage{graphicx}
\usepackage{subfigure}
\usepackage{xcolor}
\usepackage{enumerate}
\usepackage{braket}
\usepackage{amsmath, bm}

\usepackage{color, soul}
\definecolor{darkblue}{rgb}{0,0,0.5}
\setulcolor{darkblue}

\begin{document}

%\title{Density functional static dipole polarizability and first-hyperpolarizability of diatomic molecules
%in Cartesian Coordinate grid}
\title{Density functional electric response properties of molecules in Cartesian grid}

\author{Abhisek Ghosal, Tanmay Mandal}
\author{Amlan K.~Roy}
\altaffiliation{Email: akroy@iiserkol.ac.in, akroy6k@gmail.com.}                                           
\affiliation{Department of Chemical Sciences\\
Indian Institute of Science Education and Research (IISER) Kolkata \\  
Nadia, Mohanpur-741246, WB, India}

\begin{abstract}
Within the finite-field Kohn-Sham framework, static electric response properties of diatomic molecules are presented. 
The electronic energy, dipole moment ({\boldmath$\mu$}), static dipole polarizability ({\boldmath$\alpha$}) and 
first-hyperpolarizability ({\boldmath$\beta$}) are calculated through a pseudopotential-DFT implementation in Cartesian 
coordinate grid, developed in our 
laboratory earlier. We engage the Labello-Ferreira-Kurtz (LFK) basis set; while four local and non-local 
exchange-correlation (LDA, BLYP, PBE and LBVWN) functionals have been adopted. A detailed analysis of \emph{grid 
convergence} and its impact on obtained results, is presented. In each case the \emph{electric field optimization} 
was carefully monitored through a recently prescribed technique. For all three molecules (HCl, HBr, HI) considered, the 
agreement of all these quantities with widely successful and popular atom-centered-grid procedure, is excellent. To 
assess the efficacy and feasibility, companion calculations are performed for these on a representative molecule (HCl) 
at distorted geometries, far from equilibrium. Wherever possible, relevant comparison is made with available 
\emph{all-electron} data and experimental results. This demonstrates that Cartesian grid provides accurate, reliable 
results for such properties of many-electron systems within pseudopotential representation. 

%\vspace{5mm}
%{\bf PACS Numbers:} 03.65-w, 03.65Ca, 03.65Ta, 03.65.Ge, 03.67-a

\vspace{5mm}
{\bf Keywords:} Density functional theory, electric response, dipole moment, polarizability, first-hyperpolarizability,  
cartesian grid. 

\end{abstract}
\maketitle

%123456789012345678901234567890123456789012345678901234567890123456789012345678901234567890123456789012345678901234567890
\section{Introduction}
Density functional theory (DFT) \citep{hohenberg64,kohn65} has been found to be an indispensable tool for 
determining structure and properties of many-electron systems like atoms, molecules, clusters, 
nano-materials and periodic systems over the past three decades \cite{parr89, joubert98, gidopoulos03, cohen11, 
engel11, burke12, becke14, jones15}. The theory is exact, \emph{in principle} and has the ability to capture 
many-body correlation effects at a computational cost comparable to Hartree-Fock theory. Naturally, it 
has been widely applied in solid-state physics and material chemistry for understanding the physio-chemical 
phenomena. The linear and non-linear electric properties such as dipole moment, polarizability,  
hyperpolarizability are important for many applications, e.g., the development of nonlinear optical materials 
\cite{munn01}, Raman and infrared spectroscopy \cite{tanimura09}, structural identifications of atomic clusters 
\citep{wang09}, separations of molecular isomers \citep{kinnibrugh06}, etc. Throughout the past several decades, a lot 
of theoretical developments have taken place to determine the electric response properties of atoms and 
molecules using DFT as well as different \emph{ab initio} methods. Several excellent reviews are available; 
here we refer to a selective set \citep{pugh99,champagne10,helgaker12}.

The above-mentioned properties may be computed using a number of different theoretical approaches within the Kohn-Sham 
(KS) DFT rubric. Broadly speaking, there exists two distinct routes. In the first case, one obtains the response of 
density matrix analytically by solving a set of coupled perturbed KS (CPKS) equations \citep{fournier90,colwell93,
komornicki93,lee94}. \textcolor{red}{The atomic-orbital basis along with the iterative nature of this procedure makes it usually somehow
expensive in terms of computational overhead for larger systems. Moreover, this also requires information about 
analytical gradients or excited states. Two major prescriptions have been put 
forth in the literature, namely (i) a non-iterative approximation to CPKS \citep{sophy03,sophy08} and (ii) an 
auxiliary density perturbation theory \citep{moreno08,carmona12}.} The former is semi-numerical in nature where 
the derivative of KS matrix is estimated using finite-field (FF) approximation from which the response density matrix is 
calculated. The latter brings in a set of auxiliary functions to express the approximated density to calculate the 
perturbed density matrix. Further, time-dependent density functional theory (TDDFT) was invoked in the 
same spirit as CPKS method for static as well as dynamic (frequency dependent) electric responses of ground, and 
subsequently excited states of molecule \citep{casida95,gorling99,hu02}. The advantage lies in the fact that it has the 
ability to determine nonlinear optical properties such as, frequency-dependent hyperpolarizabilities, second harmonic 
generation, and etc \cite{gisbergen98}. Besides these methods, which involve variations of density matrix with respect to 
electric field, there exists another attractive approach, called perturbative sum-over states expression over all 
dipole-allowed electronic transitions. Here, time-independent/time-dependent perturbation theory is used to identify 
the expressions for static/dynamic response functions respectively, by expanding the energy in order of external 
perturbations \citep{orr71,bishop94}.

\textcolor{red}{Another important direction is the FF method, which is a common technique in quantum chemistry for 
calculating these properties, as reflected from a large volume of works recorded \citep{jasien90,guan93,calaminici98,
kamada00, bulat05,touma11}. Its real advantage lies in the ease of implementation over the analytical response theory.}
In general, the response properties are very sensitive 
towards basis set, electron correlation, relativistic effects and vibrational structure, in case of molecules. Note that, as the 
response largely results from valence electrons, sufficient diffuse functions must be included in basis set. A very careful 
consideration of polarization functions is an important requisite \citep{miadokova97} to obtain accurate values. Furthermore, 
one could use pseudopotential approximation to reduce the computational burden substantially instead of full calculations. 
In that case, their determination depends on the accuracy and transferability of pseudopotential \citep{filippetti95}. Besides, 
valence basis sets which are used for pseudopotential calculations, also suffer from similar shortcomings as standard 
all-electron basis sets \citep{labello05}. Fortunately, in many cases response of core electrons is small compared to that 
of valence electrons, which makes the calculated results acceptable and realistic. In yet another effort, 
basis-set free methods offered a practical alternative, as they do not suffer from the incompleteness of basis set. For example, 
the real-space approach of \citep{vasiliev10} provides quite comparable results for atomic and molecular 
polarizabilities within pseudopotential FF DFT framework. Another interesting development for polarizability was put forth 
by Talman \citep{talman12}, by advocating the use of Sternheimer method within the unoccupied HF and coupled perturbed 
HF approximations. 

In this communication, we report the implementation of FF method within the frame work of LCAO-MO pseudopotential KS 
formalism in Cartesian Coordinate grid (CCG) \citep{roy08,roy08a,roy10,roy11,ghosal16}. The scope and applicability of the 
method was demonstrated for a decent number of atoms/molecules for various exchange-correlation (XC) functionals in terms 
of energy, ionization energy, atomization energy, orbital energy and potential energy curves. Some preliminary results
on non-uniform grid was published as well \cite{ghosal16}. Here we investigate molecular properties, namely, 
dipole moment ({\boldmath$\mu$}), dipole polarizability ({\boldmath$\alpha$}) and first-hyperpolarizability 
({\boldmath$\beta$}), 
computed explicitly including the static electric field contribution in the time-independent KS Hamiltonian. These are 
done for three molecules (HCl, HBr, HI), taking four XC functional, \emph{viz.}, (i) local density approximation (LDA) with 
the Vosko-Wilk-Nusair (VWN) \citep{vosko80} correlation (ii) Becke \cite{becke88} exchange and Lee-Yang-Parr (LYP) 
\cite{lee88} correlation--abbreviated as BLYP (iii) Perdew-Burke-Ernzerhof (PBE) \citep{perdew96} functional (iv) 
asymptotically corrected Leeuwen-Baerends (LB94) \citep{leeuwen94} exchange plus VWN correlation--abbreviated as LBVWN. The 
last one is considered because its superior asymptotic 
long-range behavior ensures the ionization energies (obtained through HOMO) as well as higher-lying states, quite 
satisfactorily. Throughout our presentation, Labello-Ferreira-Kurtz (LFK) basis set \cite{labello05} was used, which 
appears to be a quite appropriate in the literature for such pseudopotential studies. In each case, both spatial and 
field grid optimization was performed quite carefully. We also thoroughly examine the role of XC functionals to determine 
the optimal FF strength in case of a representative test molecule (HCl). The computed quantities are compared with 
available theoretical data in order to evaluate the effectiveness of CCG in this context. Additionally, the average 
polarizability for these molecules is also reported to facilitate comparison with existing
experimental and theoretical results. Next, $\bm{\alpha}$ is probed in our test molecule HCl, at different internuclear 
separations ($R$), for which static correlation plays a crucial role. The suitability and applicability of our results 
is then correlated with respective \emph{all-electron} calculations using Sadlej \citep{sadlej92} basis set, in the range 
of $R$ covered in this work.

\section{Method of Calculation}
The single-particle KS equation of a many-electron system under the influence of pseudopotential can be written as 
(atomic unit employed unless stated otherwise), 
\begin{equation}
\bigg[ -\frac{1}{2} \nabla^2 + v^{p}_{ion}(\rvec) + v_{h}[\rho(\rvec)] + v_{xc}[\rho(\rvec)] \bigg]
 \psi_i(\rvec) = \epsilon_i \psi_i(\rvec),
\end{equation}
where $v^{p}_{ion}$ denotes the ionic pseudopotential (denoted by a ``p" superscript), expressed as, 
\begin{equation}
v^p_{ion}(\rvec) = \sum_{\Rvec_a} v^{p}_{ion,a} (\rvec-\Rvec_a).
\end{equation}
Here $v^{p}_{ion,a}(\rvec)$ represents the ion-core pseudopotential associated with atom A, located at $\Rvec_a$. The 
classical Coulomb (Hartree) potential, $v_{h}[\rho(\rvec)]$ describes the electrostatic interaction among the valence 
electrons whereas $v_{xc}[\rho(\rvec)]$ signifies XC potential, the non-classical part of effective Hamiltonian. The 
single-particle charge density is then given by, 
\begin{equation}
\rho(\rvec) = \rho^{\alpha}(\rvec) + \rho^{\beta}(\rvec) = \sum_i f^{\alpha}_{i} |\psi^{\alpha}_i(\rvec)|^2 + 
 \sum_i f^{\beta}_{i} |\psi^{\beta}_i(\rvec)|^2,
\end{equation}
where $\{ \psi^{\sigma}_{i},\sigma=\alpha \quad \mathrm{or} \quad  \beta \}$ corresponds to a set of $N$ occupied 
orthonormal spin molecular orbitals (MO) and $f_{i}^{\sigma}$s denote occupation numbers related to \textit{i}th 
spin-MO. Alternatively, in terms of the atom-centered basis functions $\{\chi_{\mu} (\rvec)\}$, one can write,  
\begin{equation}
\rho(\rvec) = \sum_{\mu,\nu}^{\mathrm{N_{c}}} P_{\mu \nu }\chi_{\mu}^{\star}(\rvec) \chi_{\nu}(\rvec),  
\end{equation}
with
\begin{equation}
P_{\mu \nu} = P_{\mu \nu}^{\alpha} + P_{\mu \nu}^{\beta}, \quad  \quad  \quad \ \ \ \ \ \ P_{\mu \nu}^{\sigma} = 
\sum_i f_{i}^{\sigma} C_{\mu i}^{\star \sigma} C_{\nu i}^{\sigma},
\end{equation}
where $P_{\mu \nu}$ denotes an element of one-body density matrix $\Pvec$ while $\mathrm{N_{c}}$ corresponds to the 
total number of contracted basis functions. In LCAO-MO approximation, the coefficients for expansion of spin-MOs 
satisfy a set of equations analogous to that in Hartree-Fock theory,  
\begin{equation}
\sum_{\nu} F_{\mu \nu}^{\sigma} C_{\nu i}^{\sigma} = \epsilon_{i}^{\sigma} \sum_{\nu}^{\sigma} S_{\mu \nu} 
C_{\nu i}^{\sigma},
\end{equation}
with the orthonormality condition, $(\Cvec^{\sigma})^{\dagger} \Svec \Cvec^{\sigma} = \Ivec$. Here $\Cvec^{\sigma}$ 
contains the respective spin-MO coefficients $\{C_{\nu i}^{\sigma}\}$, $\Svec$ is the overlap matrix corresponding to 
elements $S_{\mu \nu}$, $\bm{\epsilon}^{\sigma}$ refers to diagonal matrix of respective spin-MO eigenvalues 
$\{\epsilon_{i}^{\sigma}\}$, while $F_{\mu \nu}^{\sigma}$ is an element of KS-spin matrix conveniently partitioned as, 
\begin{equation}
F_{\mu \nu}^{\sigma} = H_{\mu \nu}^{\mathrm{core}} + J_{\mu \nu} + F_{\mu \nu}^{xc\sigma}.
\end{equation} 
In this equation, $H_{\mu \nu}^{\mathrm{core}}$ contains all one-electron contributions including kinetic energy, 
nuclear-electron attraction and pseudopotential matrix elements. All one-electron integrals are generated by standard 
recursion relations \citep{obara86} using Cartesian Gaussian-type orbitals as primitives basis functions. We employ 
the angular-momentum dependent pseudopotential form as proposed by \citep{stevens84,stevens92}, 
whereas $J_{\mu \nu}$ term signifies contribution from Hartree potential and the last (XC) term arises from non-classical 
effects. 

Now all the relevant quantities like basis functions, electron densities, MOs as well as various two-electron 
potentials are directly set up on the $3$D CCG simulating a cubic box,
\begin{eqnarray}
r_{i}=r_{0}+(i-1)h_{r}, \quad i=1,2,3,....,N_{r}~, \quad r_{0}=-\frac{N_{r}h_{r}}{2}, \quad  r \in \{ x,y,z \},
\end{eqnarray}
where $h_{r}$ denotes grid spacing along each directions and $N_x, N_y, N_z$ signify total number of grid points 
along $x, y, z$ directions respectively. In case of non-uniform grid, we usually vary $N_{x}, N_y, N_{z}$ independently 
keeping the value of $h_{r}$ fixed. Thus, electron density $\rho(\rvec)$ in this grid may be simply written as 
(``g" implies the discretized grid),
\begin{equation}
\rho(\rvec_g) = \sum_{\mu,\nu} P_{\mu \nu } \chi_{\mu}(\rvec_g) \chi_{\nu}(\rvec_g). 
\end{equation}

A major concern in grid-based approach constitutes an accurate estimation of classical electrostatic potential. Here, 
we invoke the conventional Fourier convolution method \citep{martyna99,minary02}. It has been well documented 
earlier \citep{roy08,roy08a,roy10,roy11,ghosal16}; so here only the essential details are summarized. In the end, the 
classical Coulomb potential is calculated from the following, 
\begin{equation}
v_h(\rvec_g) = FFT^{-1} \{ v_{h}(\kvec_g) \rho(\kvec_g) \} \quad \mathrm{and} \quad \rho(\kvec_g) = 
FFT\{\rho(\rvec_g) \}.
\end{equation}
The quantities $v_{h}(\kvec_g), \rho(\kvec_g)$ stand for Fourier integrals of Coulomb interaction kernel and 
density respectively. The electron density in $k$ space, $\rho(\kvec_g)$ can be obtained easily using discrete Fourier 
transform of respective real-space values. Thus the real crux of our problem is calculation of Coulomb interaction kernel, 
which has singularity at origin. In order to overcome this, we exploit an Ewald summation-type approach 
\citep{chang12}, expanding the Coulomb kernel into long- and short-range components,
\begin{equation}
v_{h}(\rvec_g)=\frac{ \mathrm{erf}(\zeta r)}{r}+\frac{ \mathrm{erfc}(\zeta r)}{r} \equiv 
v_{h_{\mathrm{long}}}(\rvec_g)+v_{h_{\mathrm{short}}}(\rvec_g),
\end{equation}
where erf(x) and erfc(x) denote error function and its complement respectively. Fourier transform of short-range 
part can be treated analytically whereas the long-range portion needs to be computed directly from FFT of corresponding 
real-space values. A convergence parameter $\zeta$ is used to adjust the range of $v_{h_{\mathrm{short}}}(\rvec_g)$, 
such that the error is minimized. 

A very crucial step in DFT calculations is choice of an appropriate XC functional. While the \emph{exact} form 
remains elusive as yet, highly accurate functionals have been reported including so-called local, ``non-local" (gradient 
and Laplacian-dependent) and hybrid ones. Finally, all the two-electron KS matrix elements are calculated directly 
through numerical integration on the grid as ($v_{hxc}$ refers to Hartree and XC potential combined), 
\begin{equation}
 \langle \chi_{\mu}(\rvec_g)|v_{hxc}(\rvec_g)|\chi_{\nu}(\rvec_g) \rangle = h_x h_y h_z \sum_g
 \chi_{\mu}(\rvec_g) v_{hxc}(\rvec_g) \chi_{\nu}(\rvec_g).
 \end{equation}

The response properties of a many-electron system can be defined by expanding field-dependent dipole moment, calculated 
from the field-induced charge distribution, as a power series in the external electric field $\Fvec$, if the field 
strength remains small, i.e.,  
\begin{equation}
\mu_i(\Fvec) = \mu_i(0) + \sum_j \alpha_{ij} F_j + \frac{1}{2} \sum_{j,k} \beta_{ijk} F_j F_k + \cdots ~.
\end{equation}
In this equation, the three terms on right-hand side characterize static dipole moment $\mu_i(0)$, dipole polarizability
$\alpha_{ij}=\frac{\partial{\mu_i}}{\partial F_j}$ and first-hyperpolarizability 
$\beta_{ijk}=\frac{\partial^2 \mu_i}{\partial F_j \partial F_k}$ \citep{mclean67}
respectively. Alternatively one may also represent it in terms of field-induced energy; and both the definitions are 
equivalent according to Hellmann-Feynman theorem \citep{feynman39}. The components of $\bm{\alpha}$ and $\bm{\beta}$ 
can be deduced from the following well-known finite-difference formulas \citep{smith78} as,
\begin{equation}
\begin{rcases}
 \alpha_{ii} F_i = \frac{2}{3}\bigg[\mu_i(F_i)-\mu_i(-F_i)\bigg]-\frac{1}{2}\bigg[\mu_i(2F_i)-\mu_i(-2F_i)\bigg] \\
 \alpha_{ij} F_j = \frac{2}{3}\bigg[\mu_i(F_j)-\mu_i(-F_j)\bigg]-\frac{1}{2}\bigg[\mu_i(2F_j)-\mu_i(-2F_j)\bigg] \\
 \beta{iii} F_{i}^{2} = \frac{1}{3}\bigg[\mu_i(2F_i)+\mu_i(-2F_i)\bigg]-\frac{1}{3}\bigg[\mu_i(F_i)+\mu_i(-F_i)\bigg] \\
 \beta{ijj} F_{j}^{2} = \frac{1}{3}\bigg[\mu_i(2F_j)+\mu_i(-2F_j)\bigg]-\frac{1}{3}\bigg[\mu_i(F_j)+\mu_i(-F_j)\bigg]
 \end{rcases}
\end{equation}

Moreover, in addition to $\bm{\alpha}, \bm{\beta}$ tensors, the experimentally determined so-called average 
polarizability, defined as, 
\begin{equation}
\bar{\alpha}=\frac{1}{3}(\alpha_{zz}+\alpha_{xx}+\alpha_{yy})
\end{equation}
can also be calculated for a given species. Now in order to obtain all these above mentioned tensors from dipole moment 
of the system (expressed as a function of external electric field $\Fvec$), one needs the perturbed density matrix 
at different field strengths, which is obtained from the self-consistent solution of Eq.~(1). Hence the 
core part of the Hamiltonian (denoted by a prime) will now be modified by a field-dependent term accordingly as, 
\begin{equation}
H_{\mu \nu}^{^\prime \mathrm{core}}= H_{\mu \nu}^{\mathrm{core}} + F_i\langle \mu|\rvec|\nu \rangle, \ \ \ \ \ \ \ \ \ 
i \in \{x,y,z\}.
\end{equation}
Here $H_{\mu \nu}^{\mathrm{core}}$ refers to unperturbed core Hamiltonian described above, $F_i$ denotes \textit{i}th 
component of applied field $\Fvec$ and $\langle \mu|\rvec|\nu \rangle$ gives the dipole moment integral corresponding to 
length vector $\rvec$. All two-body matrix elements of KS matrix will remain intact during FF calculations. 
Finally, dipole moment of a molecule can be expressed as below, 
\begin{equation}
\bm{\mu} \equiv \bm{\mu}_{el}  = \sum_{\mu \nu} P_{\mu \nu} \langle \mu|\rvec|\nu \rangle + 
\sum_{a} Z_{a} \Rvec_{a},
\end{equation}
where $Z_{a}$ and $\Rvec_a$ are nuclear charge and position of atom ``a", respectively.

\section{Computational aspects}
This section provides some of the computational details in our current calculation. Methods like the present one, require 
the choice of a suitable basis set that can correctly describe the response of electrons to an external perturbation. It 
is quite well known that it should typically contain diffuse functions to provide accurate results \citep{werner76}. While 
there exist several options for full calculations which contain $d, f$ orbitals, the choice is much 
limited for pseudopotential approximations. In the current work, we have adopted the so-called LFK basis as proposed in 
\citep{labello05}, based on a procedure to incorporate diffuse and polarization functions in familiar Sadlej basis 
set \citep{sadlej92}. It was been pointed out that response properties (mainly $\bm{\alpha}$) comparable to \emph{all-electron} 
results (using Sadlej basis) can be recovered at a lower computational cost using this basis. These are taken from 
EMSL Basis Set Library \citep{feller96}.

The pertinent molecular properties are calculated for four different XC functionals: (i) LDA--with the homogeneous electron 
gas correlation proposed by VWN (parametrization formula V of Ref.~\citep{vosko80}) (ii) 
BLYP--incorporating the popular Becke \citep{becke88a} exchange along with LYP \citep{lee88} correlation 
(iii) PBE \citep{perdew96} functional (iv) LBVWN--including LB94 exchange \citep{leeuwen94} along with VWN correlation. All 
XC functionals were adopted from density functional repository program \citep{repository} except LDA and LB94. 

The self-consistent convergence criteria imposed in this communication is slightly tighter than our earlier work 
\citep{roy08,roy08a,
roy10,roy11, ghosal16}; this is to generate a more accurate perturbed density matrix. Changes in following quantities were 
checked, \emph{viz.,} (i) orbital energy difference between two successive iterations and (ii) absolute 
deviation in a density matrix element. They both were required to remain below a certain prescribed threshold set to 
$10^{-8}$ a.u.; this ensured that the total energy maintained a convergence of at least this much. To accelerate FF 
calculations, unperturbed (field-free) density matrix was used as trial input. The value of $\zeta$ in Eq.~(11) is fixed in 
such a way that $\zeta \times L = 7$, where $L$ is chosen as the smallest side of simulating box. Such a conjecture was put 
forth in \cite{martyna99} and quite successfully implemented in CCG before \citep{ghosal16}. In order to perform discrete 
Fourier transform, standard FFTW3 package \citep{fftw05} is invoked. The resulting generalized matrix-eigenvalue problem 
is solved through standard LAPACK routine \citep{anderson99} accurately and efficiently. Relevant pseudopotential matrix 
elements in Gaussian orbitals are imported from GAMESS \citep{schmidt93} package. The scaling properties 
have been discussed earlier \cite{roy08a}.

\begingroup                      %%Table 1
\squeezetable
\begin{table}      
\caption{\label{tab:table1} Convergence of electronic energy of HCl ($R=1.275 \textup{\AA}$) in the grid ($h_r=0.3$) using LDA 
XC functional. The reference value is $-$15.433429 \cite{schmidt93}. All results are in a.u.}
\begin{ruledtabular}
\begin{tabular} {rrrlrrrrl}
 \multicolumn{4}{c}{Set~I }  & & \multicolumn{4}{c}{Set~II}      \\
\cline{1-4} \cline{6-9} 
$N_x$   &   $N_y$   &    $N_z$  &  $ \langle E \rangle $ &    & $N_x$   &   $N_y$   &    $N_z$  &  $ \langle E \rangle$ \\
\cline{1-9}
40 & 40 & 40 & $-$15.405457 &  &50 & 50 & 80 & $-$15.432989 \\
- & - & 50 & $-$15.414922 &  &54 & 54 &  - & $-$15.433262 \\
- & - & 60 & $-$15.415930 &  &58 & 58 &  - & $-$15.433361 \\
- & - & 70 & $-$15.416056 &  &62 & 62 &  - & $-$15.433401 \\
-& - & \textbf{80} & \textbf{$-$15.416061} & &66 & 66 & - & $-$15.433417 \\
- & - & 90 & $-$15.416062 &  &70 & 70 &  - & $-$15.433424 \\
- & - & 100 & $-$15.416062 & & \textbf{74} & \textbf{74} & - & \textbf{$-$15.433427} \\
50 & 50 & 50 & $-$15.432747 & &78 & 78 & - & $-$15.433428 \\
 - & - & 60 & $-$15.432955 &  &90 & 90 & 90 & $-$15.433429 \\
- & - & 70 & $-$15.432985 & &100 & 100 & 100 & $-$15.433429 \\
- & - & \textbf{80} & \textbf{$-$15.432989} & &110 & 110 & 110 & $-$15.433429 \\
- & - & 90 & $-$15.432990  & &120 & 120 & 120 & $-$15.433429 \\
\end{tabular}
\end{ruledtabular}
\end{table}
\endgroup

\section{Results and discussion}
At first, it may be worthwhile to discuss the influence of spatial grid on total energy, $\langle E \rangle$ of a 
representative system with respect to the \emph{sparsity of grid} (regulated by $N_x, N_{y}, N_{z}$) and \emph{grid spacing}
(determined by $h_r$). For this, we consider a closed shell molecule, such as hydrogen chloride (HCl) at the 
experimental bond length of $1.275$\textup{\AA} along $z$ axis, taken from NIST database \citep{johnson16} as test 
case. The formal convergence and stability of the grid is illustrated in Table~I for LDA XC functional at a grid spacing of
$h_{r} = 0.3$. As in \citep{ghosal16}, we first vary $N_z$, the number of grid points along internuclear axis, keeping the 
same along $xy$ plane static at certain reasonable value, say $N_x=N_y=40$. A glance at this table shows that as $N_z$ is 
gradually increased from 40 to 90 with an increment of 10, there is a smooth convergence in energy at around $N_z=80$ 
with a difference in total energy (or more specifically grid accuracy) of about $5 \times 10^{-6}$ a.u. between two 
successive steps. In the beginning, when $N_z$ moves from 40 to 50 to 60, one notices quite dramatic improvement in energy; 
but after that the changes are relatively less until eventually reaching convergence for $N_z$ at around 80. Now, applying 
the same procedure for fixed values of $(N_{x}, N_{y})$, say at 50, offers an energy value of $-$15.432989, converging 
again in the same neighborhood of $N_{z} \approx 80$ with grid accuracy $ 5 \times 10^{-6}$ a.u. This trend 
is maintained for other $N_x, N_y$ pairs; which is not presented here to save space. It is clear from Set~I that for each such 
$(N_{x}$, $N_{y})$ pair and a particular grid accuracy, energy attains convergence for nearly the same value of $N_z$, which 
in this case happens to be around 80. Then in Set~II in right hand side, we vary $N_{x}, N_{y}$ along $xy$ plane keeping $N_z$ 
fixed at $80$. It is apparent from Set~II that the convergence in energy takes place at $N_{x}=N_{y}=74$ with same 
grid accuracy of Set I. As a further check on the numerical stability in our solution, several additional calculations were 
performed in much extended grids (last five entries); as anticipated energy remains stable in all such grids. Besides, it 
is also verified that different increments in $N_x, N_y, N_z$ do not bring any significant change in the optimal number of 
grid points as estimated above. Repeating same steps for other functionals gave only slightly different $(N_x, N_y, N_z)$ 
triplet for desired energy convergence, which is detailed as below. 

\begingroup   %%Table 2 
\squeezetable
\begin{table}      
\caption{\label{tab:table2} Optimal sparsity in grid, in case of HCl for four different XC functionals. The reference 
values \cite{schmidt93} are $-$15.43343, $-$15.47384, $-$15.50509, for LDA, BLYP, PBE respectively. Here 
$h_r \equiv h_x=h_y=h_z$, and $R=1.275 \textup{\AA}$. All results in a.u. }
\begin{ruledtabular}
\begin{tabular} {lcccccccccccccccc}
 & \multicolumn{4}{c}{LDA}  & \multicolumn{4}{c}{BLYP} & \multicolumn{4}{c}{PBE} & \multicolumn{4}{c}{LBVWN}    \\
\cline{2-5} \cline{6-9} \cline{10-13} \cline{14-17}
$h_{r}$ & $N_{x}$ & $N_{y}$ & $N_{z}$ & $ \langle E \rangle $ & $N_{x}$ & $N_{y}$ & $N_{z}$ & $ \langle E \rangle$ & 
$N_{x}$ & $N_{y}$ & $N_{z}$ 
& $ \langle E \rangle$ & $N_{x}$ & $N_{y}$ & $N_{z}$ & $ \langle E \rangle$ \\
\cline{1-17}
0.3 & 74 & 74 & 80 & $-$15.43343 & 76 & 76 & 80 & $-$15.47383 & 76 & 76 & 80 & $-$15.50513 & 68 & 68 & 80 & $-$15.43287 \\
0.2 & 106 & 106 & 120 & $-$15.43343 & 110 & 110 & 120 & $-$15.47383 & 108 & 108  & 120 & $-$15.50509 & 98 &  98 &  110 & 
$-$15.43287 \\
0.1 & 196 & 196 & 230 & $-$15.43342 & 200 & 200 & 230 & $-$15.47383 & 198 & 198 & 210 & $-$15.50508 & 184 & 184 & 210 & 
$-$15.43286 \\
\end{tabular}
\end{ruledtabular}
\end{table}
\endgroup

Now the effect of \emph{grid spacing} on energy convergence is analyzed for three different $h_{r}$ 
($0.3$, $0.2$ and $0.1$) for all four XC functionals, namely LDA, BLYP, PBE, LBVWN. This is accomplished by 
following the simple grid optimization strategy as delineated above, maintaining a grid accuracy of $5 \times 10^{-6}$ a.u., 
all throughout. The final converged 
energies with respect to various spacings for these functionals are offered in Table~II. Clearly for a fixed $h_r$, 
the optimum $N_x, N_y, N_z$ only marginally vary from functional to functional. Also it is evident that to produce
similar-quality results in a dense grid (smaller $h_r$) typically requires larger numbers of grid points. 
The above discussion thus suggests that a $h_{r}$ of $0.3$ and an optimal grid of ($74,74,80$) is sufficiently good for all 
practical purposes in our test molecule.  

Next, we move towards the electric response properties mainly $\bm{\alpha}$ and $\bm{\beta}$ tensors. It is now well 
established that the FF procedure significantly depends on choice of an appropriate field strength and its distribution. Two 
opposite and counter-intuitive effects govern the overall behavior of response, and they both should be satisfied at same time. First, 
the field must be sufficiently large so that it can overcome the finite-precision artifacts when one numerically differentiates
the dipole moment with respect to electric field. On the other hand, it must also be small enough so that the 
contributions from remaining higher-order derivatives become negligible. In order to optimize the above mentioned 
parameters, Taylor, polynomial or rational function-based FF methods \citep{mohammed13, mohammed17, patel17} have been implemented 
quite successfully in the literature. It is also well documented that the effect of field is rather quite delicate, 
especially in case of higher-order derivatives, such as $\bm{\beta}$ and $\bm{\gamma}$. The problem is critical, for the 
numerical stability of the latter derivatives satisfies a rather \emph{narrow} range of suitable field strengths. So our plan is 
to vary electric field for a given system to determine $F_{opt}$ such that this can be used in general if possible, irrespective 
of the XC functional involved. Towards this direction, we have employed a fine field as prescribed in \citep{mohammed17}; 
accordingly the calculations are performed at discrete field strengths as given by the following equation, 
\begin{equation}
F_{n} = F_{0} \times 2^{\frac{3\mathrm{n}}{100}}
\end{equation}
with $n$ ranging from 0-160, and $F_{0} = 0.0005$ a.u. This yields a maximum field strength greater than $0.01$ 
a.u. At each $F_{n}$, the required properties are calculated using Eq.~(14) for a fixed grid and basis set as mentioned above. 

\begin{figure}             %%%Fig.1
\centering
\begin{minipage}[c]{0.30\textwidth}\centering
\includegraphics[scale=0.30]{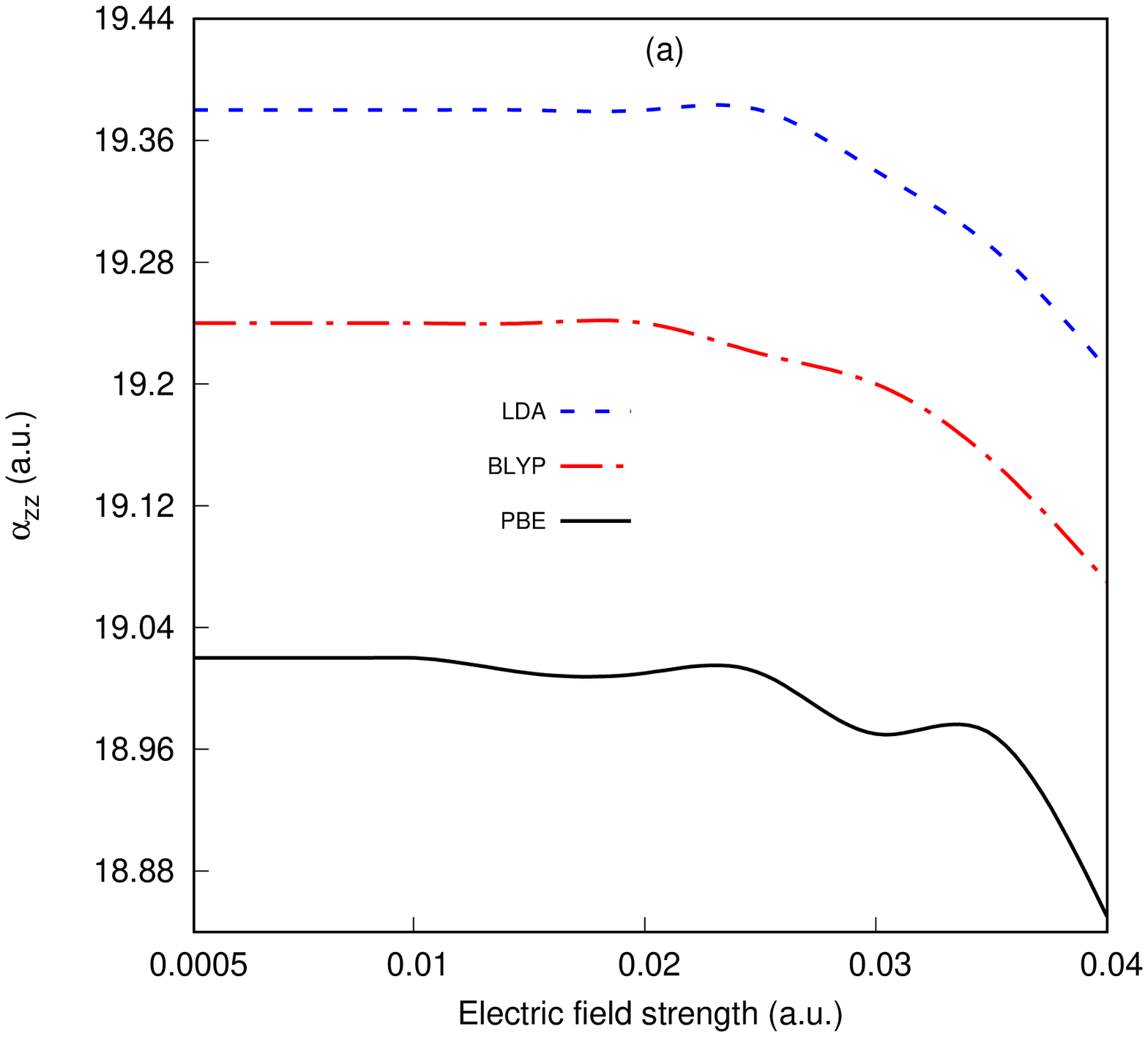}
\end{minipage}\hspace{0.1in}
\begin{minipage}[c]{0.30\textwidth}\centering
\includegraphics[scale=0.29]{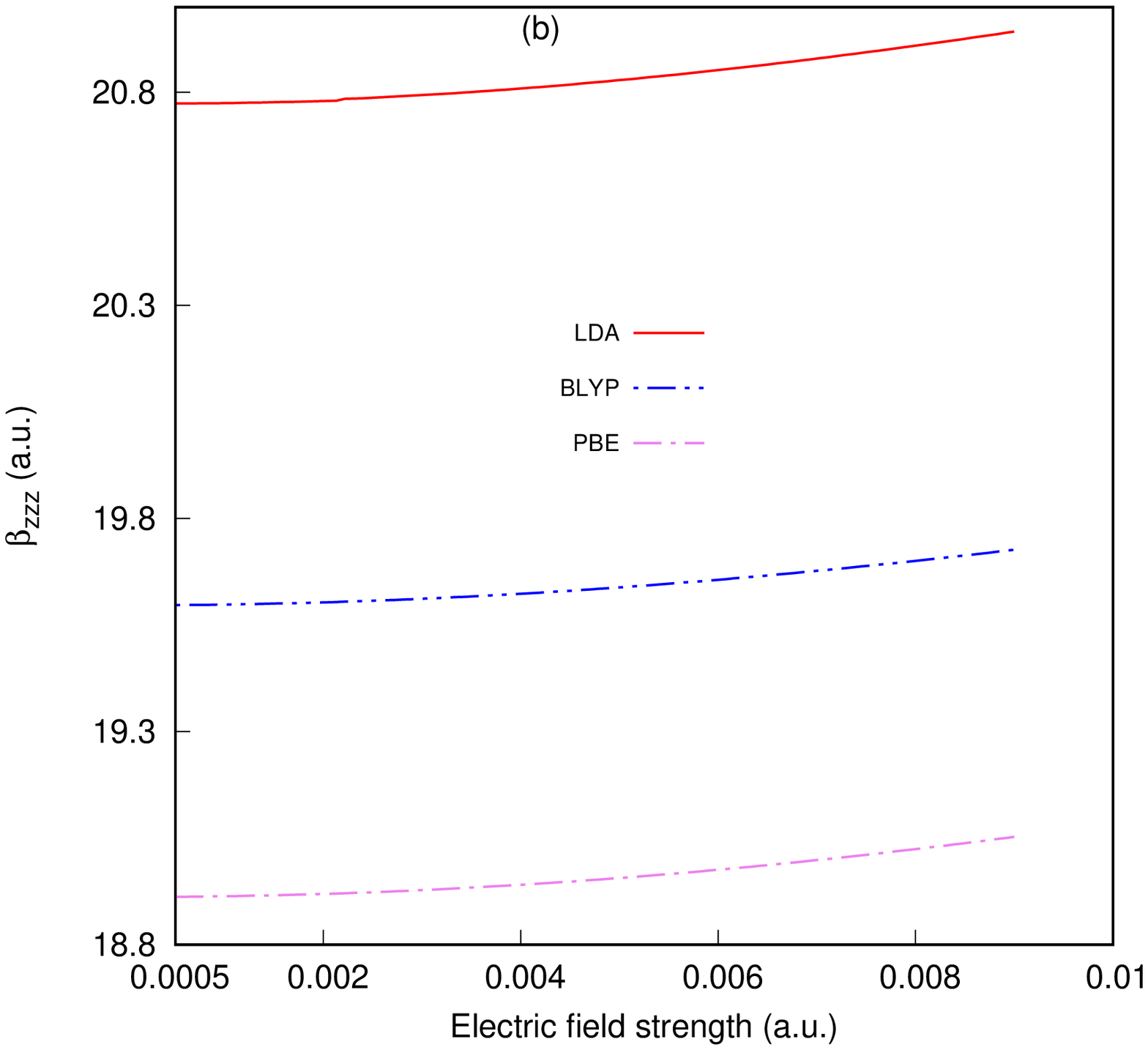}
\end{minipage}\hspace{0.1in}
\begin{minipage}[c]{0.30\textwidth}\centering
\includegraphics[scale=0.30]{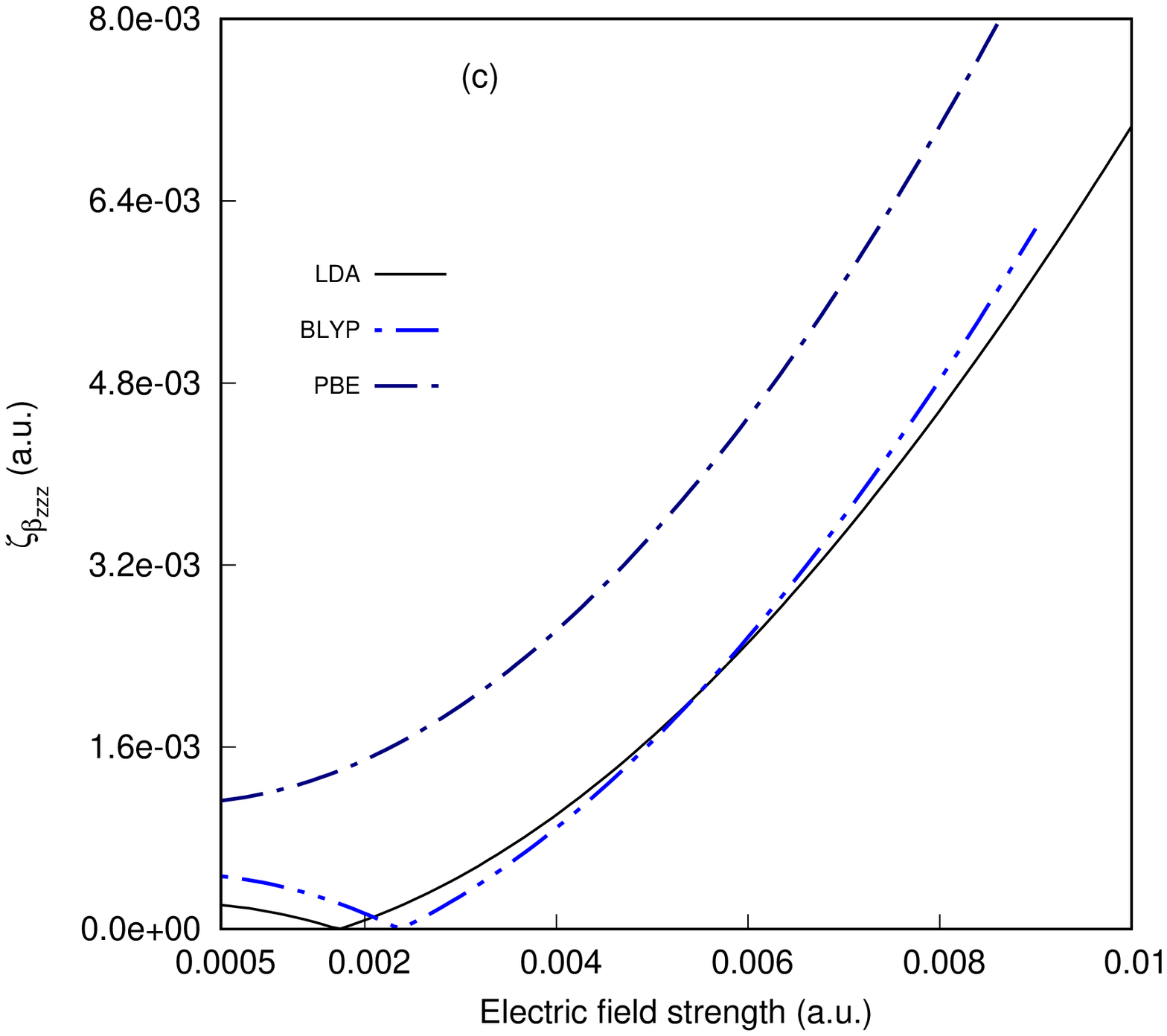}
\end{minipage}\hspace{0.1in}
\caption[optional]{Influence of electric field strength on (a) $\alpha_{zz}$, (b) $\beta_{zzz}$ and (c) $\zeta_{\beta_{zzz}}$ of HCl. }
\end{figure}

Following Buckingham \citep{buckingham67}, there are two independent components associated with 
$\bm{\alpha} \  (\alpha_{xx=yy}, \alpha_{zz})$ and $\bm{\beta} \ (\beta_{xxz=yyz}, \beta_{zzz})$ respectively, for a  
heteronuclear diatomic molecule belonging to $C_{\infty v}$ symmetry. The maximum field response towards the electron 
density is then found along $z$ direction as it is the molecular axis. So we choose only $\alpha_{zz}$ and $\beta_{zzz}$ 
among these, and show the effect of field strength on these, in Fig.~1 using three functionals (LDA, BLYP, PBE).  
From panel (a), one notices that, $\alpha_{zz}$ is practically independent of field strength for a broad region ranging 
from 0.0005 to 0.01 a.u.; this holds true all three functionals. This is in keeping with the previous work 
of \citep{calaminici98} along this direction. Therefore one is free to choose a suitable field strength in above region
for calculation of $\bm{\alpha}$ for all these functionals. In this way, a field strength of $10^{-3}$ a.u. is 
used for $\bm{\alpha}$ for HCl for all the functionals. This technique is quite general, and has been applied to other systems 
considered in this work as well. Now, a similar kind of analysis was performed for $\beta_{zzz}$ in panel (b), again
in case of HCl, for same three XC functionals as above. As expected, generally the changes with respect to field remains rather 
small in low field strength, and tends to grow in the higher range. A careful analysis reveals that, for all three 
functionals, $\beta_{zzz}$ values show a maximum deviation of up to 0.02 a.u., for a field strength ranging up to about 0.004 
a.u.

In order to analyze this effect in some detail, a field-dependent parameter 
$\zeta_{\bm{\beta}}$ is introduced which is a pointer to the relative error in $\bm{\beta}_{FF}$ with 
respect to a standard reference value, the minimum of which corresponds to $F_{opt}$ \citep{mohammed17}. This is defined as, 
\begin{equation}
\zeta_{\bm{\beta}}= \frac{\bm{\beta}_{FF}}{\bm{\beta}_{FF}^{ref}} -1.
\end{equation}
Without any loss of generality, the FF result obtained from standard GAMESS package \citep{schmidt93} is chosen as our reference. 
It may be noted that the primary objective of current methodology is to establish the suitability of this grid for molecular properties 
calculations. Accordingly, we have performed the GAMESS calculations with default options: 96 radial and 302 angular 
points for the spatial grid, and 0.001 for the field strength. In this context, it is interesting to note that, a recent study of 
grid effects (based on atom-centered grid), reported by Castet \emph{et al.} \citep{castet12}, suggested a grid of 99 radial 
and 974 angular points, to be an optimally good solution for such calculations. We have verified that for all three molecules 
considered here, the default option delivers results which are practically coincident with that from the finer grid.  
Thus the default grid serves our current purpose, as the main concern here is to validate the present grid with traditional grid.
So, taking into account this field-dependent parameter $\zeta$, the field variations for all functionals except LBVWN (for which 
the reference results are unavailable), are displayed in panel (c) of Fig.~1. One finds minima in case of both LDA and BLYP corresponding 
to the desired $F_{opt}$, but they both lie within a very narrow range of field strength from $10^{-3}$ to $3 \times 10^{-3}$ in a.u. 
However, PBE result does not show any such minimum; the plot rather gradually increases. Moreover the relative error 
($\zeta_{\bm{\beta}}$) 
in PBE plot always remains above both LDA or BLYP values for the entire field strength considered. Therefore it does not appear 
straightforward to discern an $F_{opt}$ for $\bm{\beta}$ in a given electronic system, whatever be the functional used. 
Clearly this occurs due to the narrow range of stability with respect to electric fields. On the light 
of above facts, calculation of $\bm{\beta}$ for our test molecule is performed with $F_{opt}$ corresponding to the minima in (c) 
for LDA, BLYP functionals. However, for remaining two functionals, these are chosen in the neighborhood of $1 \times 10^{-3}$ a.u.

\begingroup
\squeezetable
%\begin{table*}  
\begin{table}  
%\hfill{}   
\caption{\label{tab:table3} Static dipole moment $\mu_z$ and Finite field $\bm{\alpha}$, $\overline{\alpha}$ and $\bm{\beta}$ values (in a.u.) 
of three diatomic molecules for different XC functionals. PR implies Present Result.}
\begin{ruledtabular}
\begin{tabular} {lcccccccccccc}
 & XC & \multicolumn{2}{c}{$\mu_z$} & \multicolumn{2}{c}{$\alpha_{xx=yy}$}   &  \multicolumn{2}{c}{$\alpha_{zz}$} & 
$\overline{\alpha}^{f,g}$  
 & \multicolumn{2}{c}{$\beta_{xxz=yyz}$} &  \multicolumn{2}{c}{$\beta_{zzz}$} \\
\cline{3-4} \cline{5-6} \cline{7-8}  \cline{9-9} \cline{10-11} \cline{12-13}
 molecule & functionals  & PR & Ref. \citep{schmidt93} & PR & Ref. \citep{schmidt93} & PR & Ref. \citep{schmidt93} & PR 
& PR & Ref. \citep{schmidt93} & PR & Ref. \citep{schmidt93}\\
\cline{1-13}
% HCl$^{a,b,c}$ & LDA    & $-$0.43826 & $-$0.43826 & 18.48 & 18.48 & 19.38 & 19.38 & 18.79 & 8.26 & 8.27 & 20.77 & 20.77  \\ 
  HCl\footnotemark[1]$^,$\footnotemark[2]$^,$\footnotemark[3] & LDA    & $-$0.43826 & $-$0.43826 & 18.48 & 18.48 & 19.38 & 19.38 & 18.79 & 8.26 & 8.27 & 20.77 & 20.77  \\ 
  & BLYP    & $-$0.42337 & $-$0.42337 & 18.19 & 18.19 & 19.24 & 19.24 & 18.55 & 6.28 & 6.28 & 19.60 & 19.60  \\ 
  & PBE     & $-$0.43420 & $-$0.43425 & 18.05 & 18.04 & 19.01 & 19.01 & 18.37 & 7.19 & 7.19 & 18.91 & 19.89  \\ 
  & LBVWN   & $-$0.45357 &  -- & 15.39 & --     & 17.41 & --   & 16.07 & 3.77 & --    & 15.20 & --      \\
\cline{1-13} 
  HBr\footnotemark[4]$^,$\footnotemark[5] & LDA    & $-$0.31611 & $-$0.31612 & 25.33 & 25.32 & 26.58 & 26.58 & 25.74 & 7.27 & 7.25 & 23.13 & 23.12  \\ 
  & BLYP    & $-$0.29881 & $-$0.29882 & 24.71 & 24.70 & 26.16 & 26.16 & 25.19 & 4.16 & 4.16 & 20.90 & 20.90  \\ 
  & PBE     & $-$0.31207 & $-$0.31208 & 24.63 & 24.64 & 25.99 & 25.99 & 25.08 & 5.79 & 5.74 & 20.51 & 20.49  \\ 
  & LBVWN   & $-$0.30761 &  -- & 21.31 & --     & 24.04 & -- & 22.22 & 2.19 & --    & 15.60 & --      \\
\cline{1-13} 
  HI & LDA    & 0.18225 & 0.18226 & 37.38 & 37.37 & 39.13 & 39.13 & 37.96 & $-$3.11 & $-$3.17 & $-$16.47 & $-$16.36  \\ 
  & BLYP    & 0.16093 & 0.16103 & 36.46 & 36.48 & 38.34 & 38.33 & 37.08 &1.21 & 1.21 & $-$12.66 & $-$12.63  \\ 
  & PBE     & 0.17943 & 0.17945 & 36.35 & 36.33 & 38.18 & 38.18 & 39.24 & $-$1.84 & $-$1.73 & $-$13.25 & $-$13.19  \\ 
  & LBVWN   & 0.11947 & -- & 32.15 & --     & 35.74 & -- & 33.34 & $-$2.44 & -- & $-$9.18  & --      \\
\end{tabular}
\end{ruledtabular}
\begin{tabbing}
$^a$CAS result in taug-cc-pVTZ basis \citep{bishop99}: $\mu_z = 0.45$, $\alpha_{xx=yy} = 16.86$, 
$\alpha_{zz} = 18.52$, $\overline{\alpha} = 17.41$, $\beta_{xxz=yyz} = -0.31$, $ \beta_{zzz} = -11.32$. \\
$^b$CAS result in qaug-sadlej basis \citep{fernandez98}: $\alpha_{xx=yy} = 16.6952$, 
$\alpha_{zz} = 18.3361$, $\overline{\alpha} = 17.2422$, $\beta_{xxz=yyz} = 0.64$, $ \beta_{zzz} = 12.71$. \\
$^c$CCSD(T) result in KT1 basis \cite{maroulis98}: $\mu_z = 0.4238$, $\alpha_{xx=yy} = 16.85$, 
$\alpha_{zz} = 18.48$, $\overline{\alpha} = 17.39$, $\beta_{xxz=yyz} = -0.2$, $\beta_{zzz} = -10.7$. \\
$^d$CAS result in taug-cc-pVTZ basis \citep{bishop99}: $\mu_z = 0.36$, $\alpha_{xx=yy} = 23.52$, 
$\alpha_{zz} = 25.53$, $\overline{\alpha} = 24.19$, $\beta_{xxz=yyz} = 1.41$, $ \beta_{zzz} = -11.13$. \\
$^e$CAS result in qaug-sadlej basis \citep{fernandez98}: $\alpha_{xx=yy} = 23.4521$, 
$\alpha_{zz} = 25.1386$, $\overline{\alpha} = 24.0143$, $\beta_{xxz=yyz} = -0.81$, $ \beta_{zzz} = 11.14$. \\
$^f$The experimental $\overline{\alpha}$ of HCl from dipole (e,e) method \citep{olney97} is: 16.97. \\
$^g$The experimental $\overline{\alpha}$ of HCl, HBr and HI, from refractive index method \citep{hohm13} are: 17.40, 
23.78, 35.30.
\end{tabbing}
\end{table}
\endgroup

In order to extend the scope and applicability of current scheme, we now report the non-zero components of FF 
$\bm{\alpha}$, $\bm{\beta}$ along with $\overline{\alpha}$ and static dipole moment $\mu$, of two 
selective molecules (HBr, HI) along with the test molecule in Table III. These are provided for all four XC 
functionals. To put things in perspective, we also quote the reference values (except LBVWN) obtained from GAMESS software 
\citep{schmidt93}. Same strategy as in HCl has been followed for the other two--experimental geometries are taken from NIST 
database \citep{johnson16}, active 
grid for each molecule was optimized accordingly, keeping the molecular axis along z direction to be fixed, and then followed 
up the FF procedure as mentioned earlier. The fact that $\bm{\alpha}$ remains stable for a relatively broad range of field 
strength, enables us to fix the $F_{opt}$ safely at $10^{-3}$ for all these molecules (for all functionals). The same field 
optimization protocol as in HCl was applied for all of them to get an idea about $F_{opt}$, for $\bm{\beta}$. The maximum 
absolute deviation (MAD) in $\mu_{z}$ in HCl is around $5 \times 10^{-5}$ a.u. for PBE, whereas the same match perfectly 
with reference (in all the digits quoted) for LDA and BLYP. The same in case of HBr, HI agree quite nicely with reference--MAD 
for LDA and PBE being 0.00001 and 0.00002 a.u. respectively, while for BLYP it is 0.0001 (in HI). The $\bm{\alpha}, \bm{\beta}$ 
tensors of our calculations are also equally consistent with the reference data, the respective MAD's in 
$\alpha_{xx=yy}$ and $\alpha_{zz}$ being 0.02 and 0.01 respectively (both occur for HI). For $\beta_{xxz=yyz}$ and $\beta_{zzz}$, 
respective MAD's turn out to be 0.11 in both occasions (again for HI). We also quote some relevant theoretical 
results for HCl and HBr in the footnote, along with the methods (such as higher-order perturbation theory, MCSCF, 
CCSD(T) etc.) and basis set. Similarly, experimental values for 
$\overline {\alpha}$ are also recorded from two different kinds of experimental techniques \citep{olney97,hohm13}.  
These values contain only electronic part and neither geometry relaxation in presence of electric field, nor vibrational 
contribution are considered. It reveals an interesting fact that, 
all three traditional functionals (LDA, BLYP, PBE) overestimate both experimental results in the range of $3$-$11\%$, 
$1$-$9\%$ and $3$-$8\%$ respectively; however, LBVWN underestimates by $5$-$9\%$. Similar conclusions have also been drawn 
regarding the pattern behavior of these functionals for $\overline{\alpha}$ in \citep{cohen99,vasiliev10}, where it has been 
conjectured that, a significant improvement may be achieved for $\bm{\alpha}$ by combining LB94 potential (in asymptotic region) 
with LDA (corrected for derivative discontinuity in the bulk region) exchange suitably. This leads a more accurate representation 
of exchange which approaches the experimental results quite closely \citep{casida98,casida00} at a lower level of computational 
cost than standard XC functionals. 
 
\begingroup
\squeezetable
\begin{table}     
\caption{\label{tab:table4} Static dipole moment $\mu_z$, FF $\bm{\alpha}$, $\overline{\alpha}$ and $\bm{\beta}$ 
values (in a.u.) of HCl molecule at various distorted geometries. All quantities are in a.u. PR implies Present Result.}
\begin{ruledtabular}
\begin{tabular} {lccccccccccc}
 $R$ & XC  & \multicolumn{2}{c}{$\mu_z$} & \multicolumn{2}{c}{$\alpha_{xx=yy}$}   &  \multicolumn{2}{c}{$\alpha_{zz}$}  & 
\multicolumn{2}{c} {$\beta_{xxz=yyz}$}   &  \multicolumn{2}{c}{$\beta_{zzz}$} \\
\cline{3-4} \cline{5-6} \cline{7-8} \cline{9-10} \cline{11-12}
 & functional & PR & Ref.~\citep{schmidt93} & PR & Ref.~\citep{schmidt93}  & PR & Ref.~\citep{schmidt93}  & PR   & 
Ref.~\citep{schmidt93}  & PR & Ref.~\citep{schmidt93} \\
\cline{1-12}
  1.5      &  LDA   & $-$0.31111 & $-$0.31111 & 16.80 & 16.80 & 13.85 & 13.85 & 20.29 & 20.29 & 18.59 & 18.59 \\ 
           &  BLYP  & $-$0.31210 & $-$0.31213 & 16.69 & 16.69 & 13.64 & 13.64 & 19.20 & 19.19 & 16.55 & 16.52 \\ 
           &  PBE   & $-$0.32008 & $-$0.32013 & 16.46 & 16.46 & 13.52 & 13.52 & 19.37 & 19.35 & 17.05 & 16.89 \\ 
\cline{1-12}
  2.0      &  LDA   & $-$0.37343 & $-$0.37344 & 17.71 & 17.71 & 16.38 & 16.38 & 15.24 & 15.25 & 20.66 & 20.67 \\ 
           &  BLYP  & $-$0.36844 & $-$0.36844 & 17.50 & 17.50 & 16.20 & 16.20 & 13.68 & 13.67 & 19.14 & 19.15 \\ 
           &  PBE   & $-$0.37698 & $-$0.37703 & 17.14 & 17.14 & 15.42 & 15.42 & 15.56 & 15.56 & 18.66 & 18.63 \\ 
\cline{1-12}
  2.5      &  LDA   & $-$0.45428 & $-$0.45429 & 18.67 & 18.67 & 20.19 & 20.19 & 6.48  & 6.48  & 20.98 & 20.99 \\ 
           &  BLYP  & $-$0.43630 & $-$0.43630 & 18.36 & 18.35 & 20.06 & 20.06 & 4.39  & 4.40  & 19.84 & 19.83 \\ 
           &  PBE   & $-$0.44800 & $-$0.44804 & 18.22 & 18.21 & 19.81 & 19.81 & 5.39  & 5.39  & 19.07 & 19.07 \\ 
\cline{1-12}
  3.0      &  LDA   & $-$0.55506 & $-$0.55506 & 19.63 & 19.63 & 25.44 & 25.44 & $-$3.83  & $-$3.82  & 26.01 & 26.01 \\ 
           &  BLYP  & $-$0.51364 & $-$0.51361 & 19.22 & 19.21 & 25.42 & 25.42 & $-$6.49  & $-$6.47  & 24.69 & 24.68 \\ 
           &  PBE   & $-$0.53309 & $-$0.53312 & 19.13 & 19.13 & 25.05 & 25.05 & $-$4.98  & $-$4.99  & 23.25 & 23.26 \\ 
\end{tabular}
\end{ruledtabular}
\end{table}
\endgroup

In the last part of this study, we investigate the efficacy of CCG in determining the non-zero components of $\mu$ as well as 
$\bm{\alpha}$ and $\bm{\beta}$ tensors at different internuclear separations ($R$) in Table IV. As an illustration, once 
again, HCl has been chosen with $R$ ranging from 1.5-3.0 a.u. In general, beyond equilibrium geometry the static correlation 
becomes dominant; hence, the role of XC functional is of utmost importance. Moreover, the role of basis set is also a major 
factor in the estimation of $\bm{\alpha}$, $\bm{\beta}$ in such regions, and it will be discussed shortly in next paragraph. 
The optimal spatial mesh is determined at each $R$ using the automated grid optimization technique of Table~I. The same 
optimal field strengths of Table~III were adopted, as these do not affect the qualitative nature of present results with respect 
to $R$. All relevant quantities are recorded in Table~IV at four distinct (1.5, 2, 2,5, 3) $R$. The computed $\mu_z$ 
values are in very excellent agreement with theoretical references, for all XC functionals throughout the whole region; maximum 
discrepancy ($5 \times 10^{-5}$) occurs in case of PBE at $R=2$ a.u. A similar comparison for $\alpha_{zz}$ and 
$\alpha_{xx=yy}$ reveals that, in the former, reference results are completely reproduced by CCG, again for all functionals 
for all $R$'s considered, while in the latter case, the two results remain separated by a MAD of 0.01 in few occasions. 
Similarly, the agreement of $\bm{\beta}$ components with reference is also excellent for all $R$'s with MAD being 0.16 a.u., 
for the lone case of PBE at $R=1.5$. It may occur due to the fact that $F_{opt}$ is in general, affected by molecular size 
in case of $\bm{\beta}$, and we have not taken this into account (same $F_{opt}$ is tacitly assumed to be valid for all $R$, 
which may not hold true). A closer look at this table further reveals that there is a change in sign in $\beta_{xxz=yyz}$ on 
varying 
$R$ from 2.5 to 3 a.u., which is quite satisfactorily captured in our results. Lastly to conclude this portion, no attempt was 
made to do field optimization at each $R$. While this has little or practically no bearing on $\alpha$, the estimation of 
$F_{opt}$ may have appreciable effect in estimating $\bm{\beta}$ at different distorted geometries. 
   
\begin{figure}             %%%Fig.2
\centering
\begin{minipage}[c]{0.42\textwidth}\centering
\includegraphics[scale=0.45]{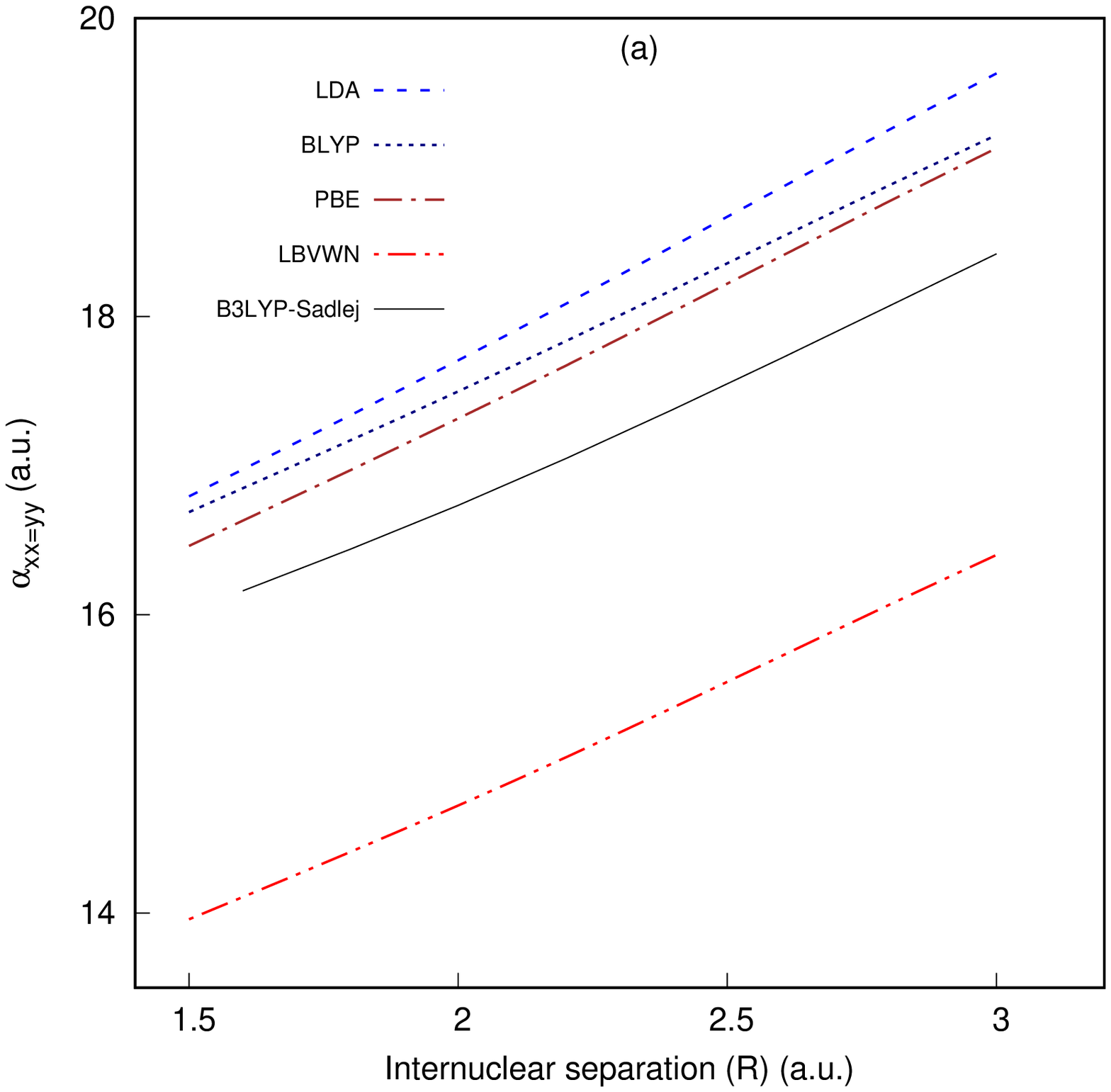}
\end{minipage}\hspace{0.8in}
\begin{minipage}[c]{0.42\textwidth}\centering
\includegraphics[scale=0.45]{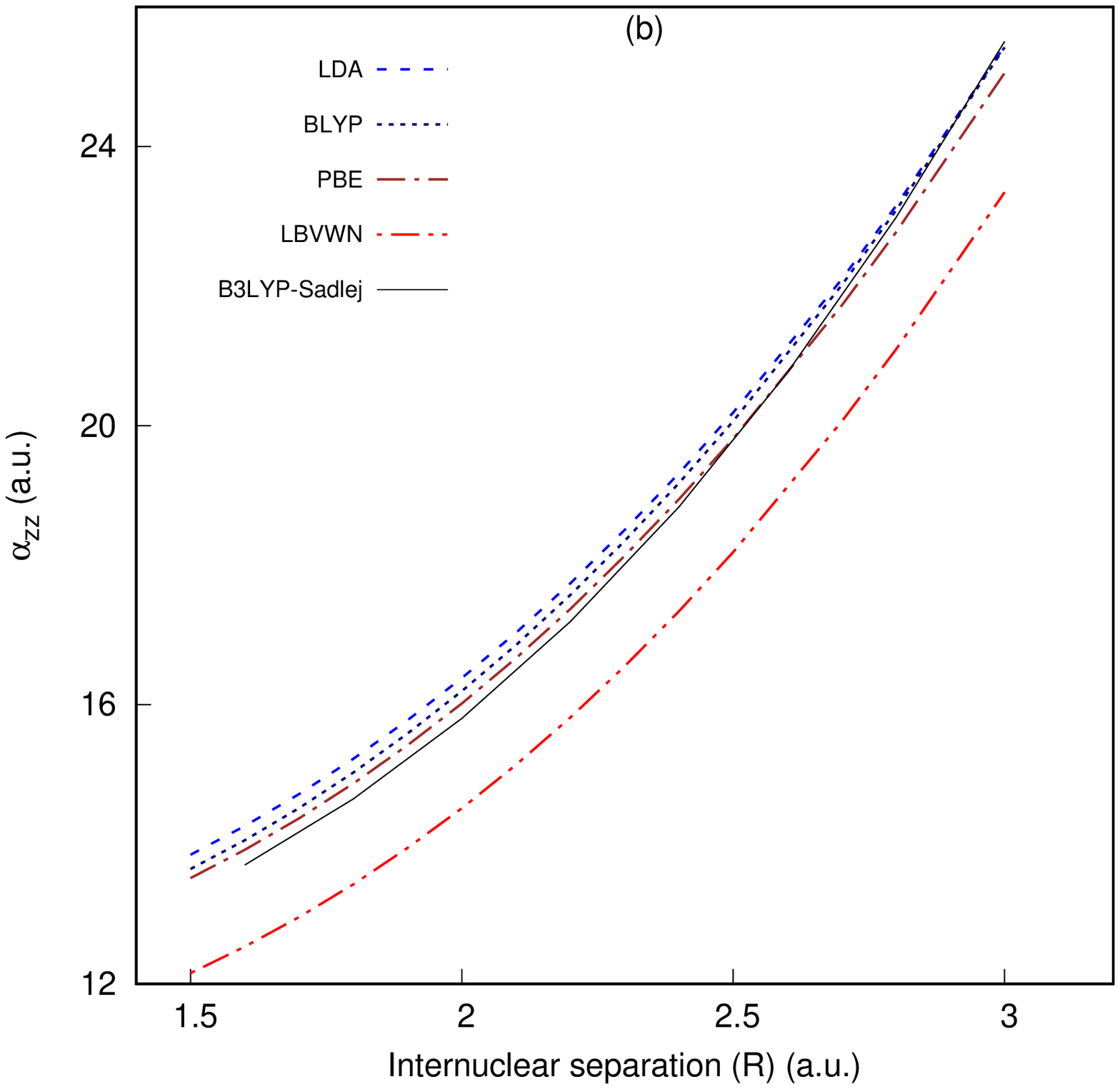}
\end{minipage}\hspace{0.001in}
\caption[optional]{Influence of internuclear separation  on (a) $\alpha_{xx=yy}$ and (b) $\alpha_{zz}$ of HCl molecule.}
\end{figure}

Next we examine the impact of XC functional and basis set on changes in $\bm{\alpha}$ at distorted geometry (of same HCl 
molecule) and compare with \emph{all-electron} results. As far as our knowledge goes, such an analysis of the performance of 
LFK basis in 
determining the response properties beyond equilibrium geometry $(R_{eq})$, has not been undertaken before. Figure~2 
portrays the calculated $\alpha_{xx=yy}$ and $\alpha_{zz}$ (segments (a), (b) respectively) with $R$, covering a broad 
region of 1.5-3 a.u., at intervals of $0.1$ a.u. The all-electron results are done using Sadlej basis \citep{sadlej92} 
and standard B3LYP functional through the GAMESS program. All the functionals reproduce the qualitative shape of 
$\alpha_{xx=yy}$ and $\alpha_{zz}$ very well for the entire range. It is noticed that in both panels, PBE results are the 
closest to Sadlej-B3LYP results around $R_{eq}$. While in panel (a) all the plots remain well separated, a distinct crossover 
is noticed in panel (b) as one moves farther beyond $R_{eq}$. On closer inspection, PBE plot in (b) tends to deviate maximum 
from the all-electron results amongst all functionals. Thus our present pseudopotential FF DFT calculation (with the aid of 
LFK basis) in CCG can produce comparable results for $\alpha_{xx=yy}$, $\alpha_{zz}$ with the more elaborate \emph{full} 
calculations, both around and far from equilibrium. 

A few remarks may now be made before passing. In this work our primary objective was to demonstrate that the real-space CCG 
coupled with FF method (for electric response calculations) could deliver accurate and physically meaningful results within 
the chemical accuracy, for diatomic molecular systems. This was mostly done through comparisons with standard-grid results. 
No effort was made to reproduce either accurate theoretical (within DFT as well as wave-function based) or experimental results, 
which may be considered in future, along with non-linear molecular systems. These would necessarily require the inclusion of 
more precise $F_{opt}$, especially for higher-order derivatives, more accurate XC functionals having correct short- as well 
as long-range properties. 

\section{Future and outlook}
We have presented a detailed study on the performance of CCG in the context of $\mu, \bm{\alpha}, \bm{\beta}$ 
of diatomic molecules, using first-principles pseudopotential DFT formalism in amalgamation with the FF method. This was 
achieved through an accurate representation of FF formalism in real-space grid. The viability and feasibility of this approach 
has been demonstrated by applying it to a set of three diatomic molecules. Four different representations of XC 
functional was invoked. The effect of spatial grid as well as optimal electric field was analyzed in detail. It is quite
gratifying that our results are in excellent agreement with those from standard program using atom-centered grid. 
In addition, for the first time, the effectiveness of pseudopotential LFK basis set (in CCG) is compared with the 
\emph{all-electron} results, far from equilibrium. Application of this approach to more chemical systems would further 
enhance its success, which may be pursued in future. To conclude, pseudopotential-CCG can offer fairly accurate and reliable 
results for electric response properties of many-electron systems. 

\section{Acknowledgement}
AG is grateful to UGC for a Senior Research Fellowship (SRF). TM acknowledges IISER Kolkata for a Junior Research 
fellowship (JRF). Financial support from DST SERB, New Delhi, India (sanction order number EMR/2014/000838) is gratefully 
acknowledged. We sincerely thank the two anonymous referees for their valuable and constructive suggestion, which have 
improved the manuscript greatly. 

\bibliography{dftbib.bib}
\bibliographystyle{unsrt}

\end{document}